# Study of dissipation processes in high-temperature superconductors of $ErBa_2Cu_3O_{7-\delta}$ system near the superconductive transition critical temperature $T_c$ in the range of possible Abrikosov vortex lattice melting.

**S. M. Ashimov and J. G. Chigvinadze**

Andronikashvili Institute of Physics, Georgian Academy of Sciences

Abrikosov vortex melting in high-temperature superconductors of $ErBa_2Cu_3O_{7-\delta}$ system is investigated using mechanical contactless method of energy dissipation.

Investigations made with a help of torsional balance [1÷5] make it possible to study phase transitions in the vortex lattice.

As it is known [1,6], the Abrikosov vortex melting takes place near $T_c$. With that the character of vortex motion essentially changes [7].

Firstly, in this temperature range where the vortex lattice transition in the liquid state takes place, its stiffness should sharply decrease it doesn't disappear at all. Consequently, vortex should comparatively freely move in a crystal lattice of high-temperature superconductor and in addition naturally the dissipation energy should essentially increase.

Also it is well-known that at low temperatures $T \ll T_c$ in a high-temperature superconductor there is a clearly formed Abrikosov vortex lattice and relaxation processes in this lattice are described by the logarithmic law [1,6].

Secondly, the relaxation processes in a vortex lattice should drastically change at its melting.

Really, at the Abrikosov vortex lattice melting in high-temperature superconductors the logarithmic relaxation law is changed on a power one with factor 2/3 [8].

So, the power law of relaxation describes a new phase where the vortex lattice stiffness should disappear or, at the least degree, sharply diminish and the vortex system should behave more similarly that of liquid.

Thirdly, approaching critical temperature $T_c$ ( near $T_c$ ), as a rule in the range of vortex lattice melting, the pinning force starts to decrease what facilitates a more free vortex motion the in temperature range far away from $T_c$.

So, at the Abrikosov vortex melting we should obtain a sharp decrease of vortex lattice dynamics in high-temperature superconductors we selected the precision contactless mechanical method of investigation of superconductor and its vortex stiffness [9, 10].

Small vibration of superconductive cylinder suspended by a thin elastic thread in a magnetic field with the strength $H>H_{c1}$ directed perpendicularly to its axis and performing axial-and-torsional vibration are observed.

With a change of vibration frequency at $H>H_{c1}$ one could make conclusion on a stiffness of vortex lattice, and a change of logarithmic decrement of damping gives information about the dissipation energy caused by the motion of Abrikosov vortices detached from the pinning-centers.

In works made by H. Keller with collaborators [11] the vortex lattice stiffness change was investigated in monocrystal samples of high-temperature superconductors at the transition of vortex matter in a new phase state where the vortex lattice melts near Tc and experiences transition in a liquid state.

Last time problems similar to those investigated by Keller with collaborators and to that experiment, the description of which is presented in this work, that is the vortex lattice dynamics study in high-temperature superconductors, are studied using monocrystal samples what simplifies otherwise a very complicated picture and makes the interpretation of results more clear and reasonable. But high-temperature superconductors of (123) system are such a fine effects as Abrikosov vortex lattice dynamics in these small monocrystals are difficult to observe. For this reason one should resort to the help of different methods of processing the experimental data to reveal different interesting peculiarities which are different to registrate due to the smallness of used samples.

Hence we decided to fabricate HTSC texturized samples with larger sizes **c**-axes of which could be directed on one's desire either along the axis of cylinder to study the AVL dynamics in the basal plane, or the c-axis could be directed to the cylinder axis to study the AVL dynamics of vortices directed along c-axis.

This arrangement of experiment gives also possibility for the investigation of Abrikosov vortex lattice dynamics on its orientation. It should be pointed that the texturized sample doesn't certainly have such ideal structure as one in the high-quality monocrystal HTSC samples.

Only that basic fact that in the basal plane of texturized sample **a** and **b** axes are disoriented shows that it is considerable inferior to the monocrystal quality, But such sample is in many respects preferable then a ceramic sample and he is intermediate between ceramics and monocrystal. In any case in the texturized samples one could study the AVL dynamics either in a basal plane or in the case when Abrikosov vortices are directed along c-axis. For the applications of texturized sample it is most attractive that we could have sample with its sizes of many times greater (in orders) then sizes of monocrystal and correspondingly, effects appearing at the AVL dynamics study are considerably greater then ones that could be expected or observed in monocrystal samples.

Taking into account all above pointed considerations we made texturized $ErBa_2Cu_3O_{7-\delta}$ samples by the standard method [12].

Cylindrical samples with the **c**-axis directed along cylinder axis and critical temperature of superconducting transition is Tc=93.5 by the standard method [12]. The cylinder is suspended by a thin elastic thread and placed in a magnetic field perpendicular to the cylinder axis creating the Abrikosov vortex lattice. By the investigation of small vibrations in these conditions we study the basal plane.

In Fig.1 the temperature dependence of the logarithmic decrement of damping at different magnetic fields applied cylindrical $ErBa_2Cu_3O_{7-\delta}$ , sample is shown. While taking each point in these figures as well as in Fig 2 and 3 temperature is constant.

As it is seen from figure near $T_c$ the damping of vibrations begins sharply increase and this increase starts at the lower temperatures the larger magnetic field is.

The beginning of increase of the vibration damping could be connected with the start of Abrikosov vortex lattice melting process during which bound and forced lines are escaped and become more mobile what apparently reveals itself in the increase of suspense system vibration damping. Further temperature increase ($T>T_c$) in the normal state the damping comes out on a plateau and in the investigated region up to T=102K the damping is not changed. With the decrease of magnetic field strength the damping $\delta_{max}$ decreases in its absolute value and displaces up to more high temperature. The decrease of damping $\delta$ at the increase of temperature, i.e. after the passing of maximum $\delta_{max}$ most probably should be connected with the transient processes of high-temperature superconductor in the normal state.

If the beginning of a sharp increase of damping $\delta$ could be connected with the start of Abrikosov $\delta$ vortex melting and it is marked as $\delta_{melt}$ then the region between $\delta_{melt}$ and $\delta_{max}$ presented in Fig.2 could be apparently marked as the region of Abrikosov vortex lattice melting.

In the Fig.3 the dependence of suspension system vibration frequency at different values of magnetic field. As it is seen from the figure in the range where an increase of damping is observed the vibration frequency decreases more sharply, then at more lower temperatures but before the start of beginning of the vortex lattice melting on the dependence $\omega(T)$ there is a cleanly expressed minimum of frequency which could be apparently be an indication of the beginning of Abrikosov vortex lattice melting. It should be also pointed that the larger the strength of applied magnetic field creating Abrikosov vortices, the clearer the above mentioned minimum is seen.

In conclusion, we suppose that the application of mechanical method gives the opportunity to study phase transitions in the vortex matter, in particular, in the experiments described in this work the Abrikosov vortex lattice melting in the basal plane was apparently observed.

Acknowledgements:
This work is made with support of International Scientific and Technology Center (ISTC) through Grant G-389.

Figure captions

Fig.1. Logarithmic decrement of damping $\delta$ dependence of sample vibrations on temperature in different magnetic fields.

Fig.2. Dependence of $\delta_{melt}$ and $\delta_{max}$ on temperature.

Fig.3. Vibration frequency of sample $\omega$ dependence on temperature.

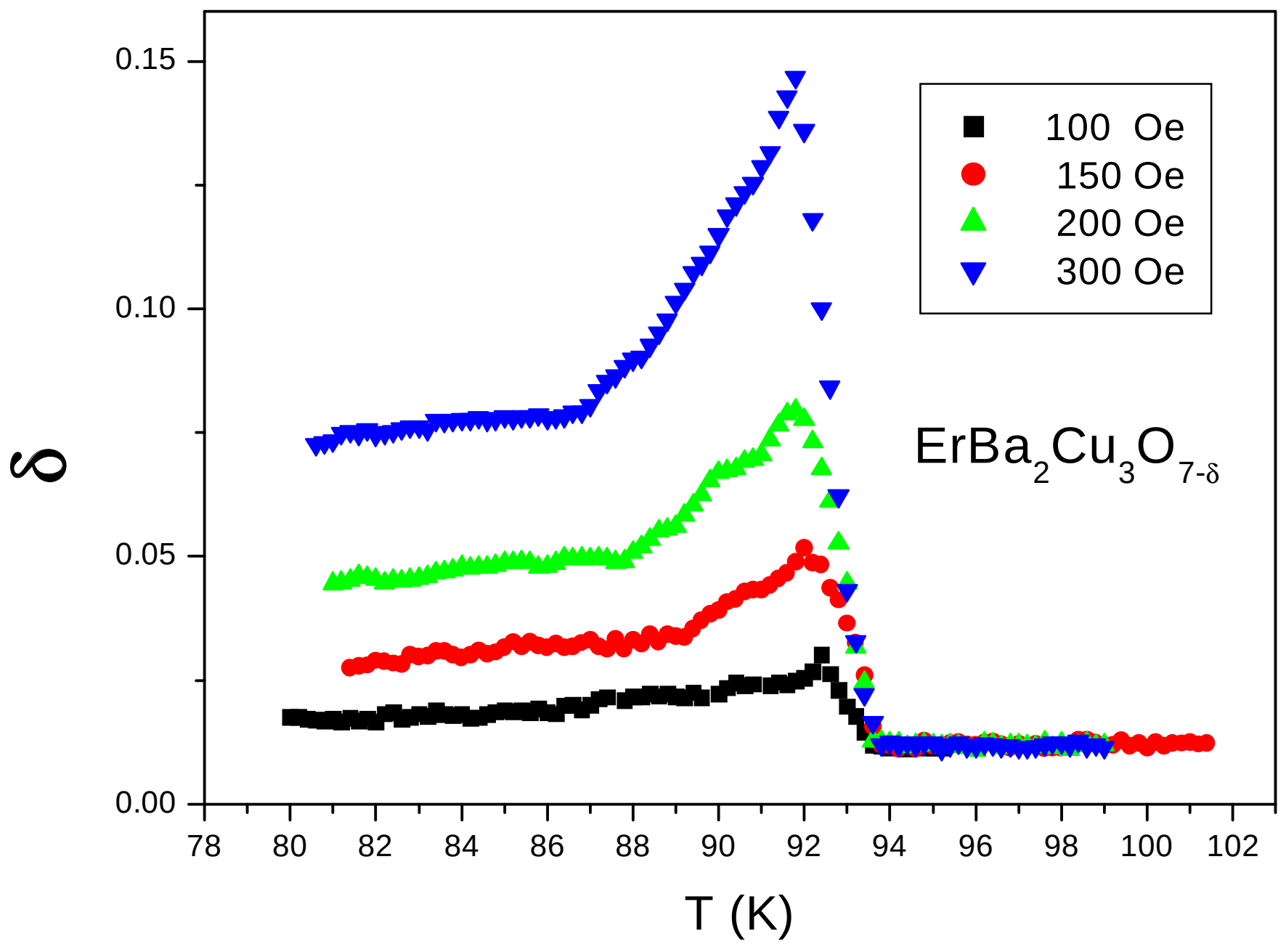

Fig.1.

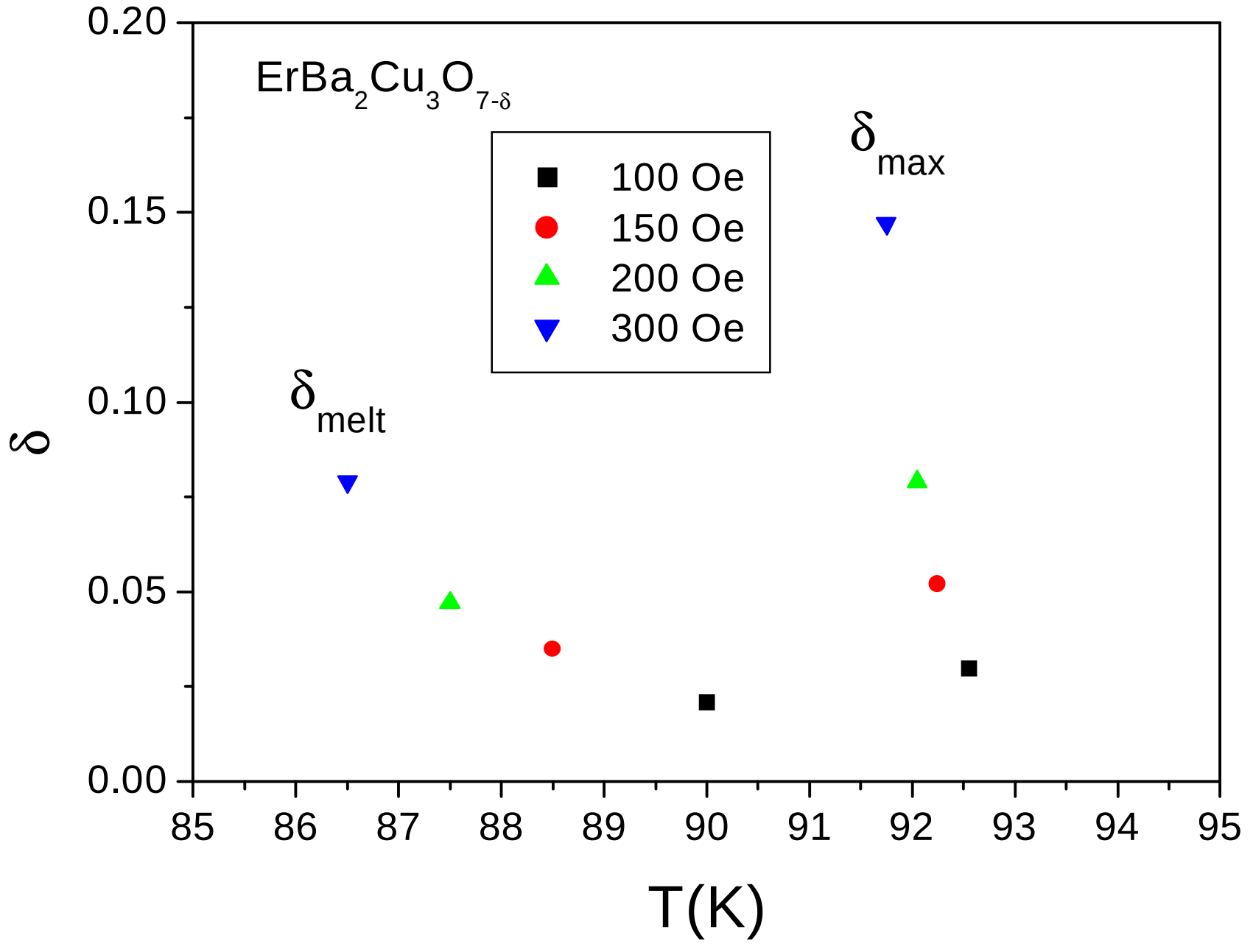

Fig.2.

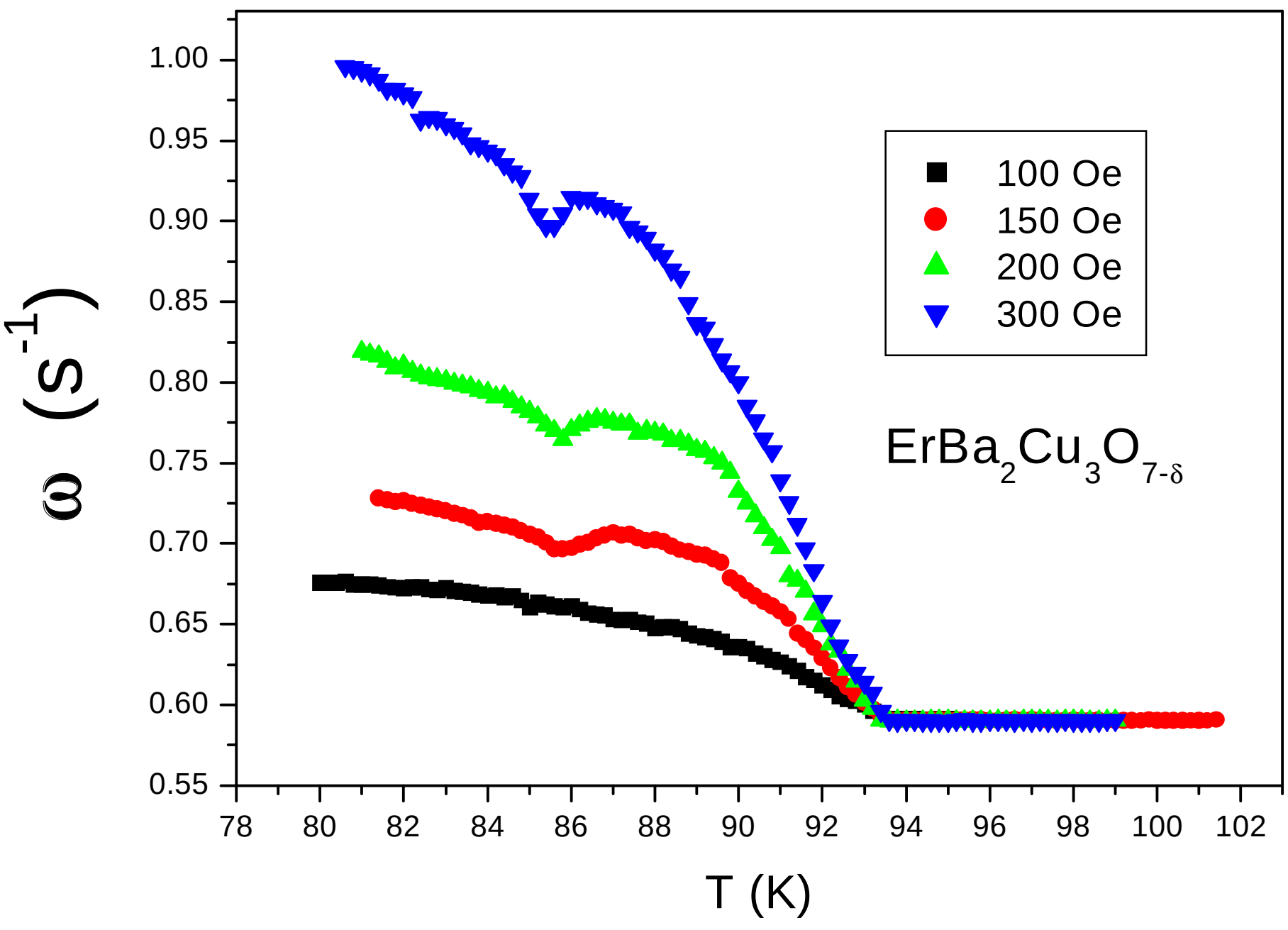

Fig.3.